\documentclass[a4paper,11pt]{article}

\usepackage{amssymb,amsmath,amsfonts,epsfig,pstricks,pst-coil,graphics,pst-node,pst-plot}
\usepackage{psfrag}
\usepackage{tikz}
\usetikzlibrary{shapes, arrows, shadows}

\def\be{\begin{equation}}
\def\ee{\end{equation}}
\def\bea{\begin{eqnarray}}
\def\eea{\end{eqnarray}}

\newcommand{\re}{\mathrm{Re\,}}

\newcommand{\cor}[1]{\left\langle #1 \right\rangle}

\newcommand{\ket}[1]{|#1\rangle}
\newcommand{\bra}[1]{\langle #1 |}
\newcommand{\ishiket}[1]{|#1\rangle\!\rangle}

\newcommand{\braneket}[1]{\|#1\rangle\!\rangle}

\newcommand{\comment}[1]{}

\newlength{\myl}

\newcommand{\dcox}{\mathsf{g}^{\!\vee}}

\let\oldsqrt\sqrt
\def\sqrt{\mathpalette\DHLhksqrt}
\def\DHLhksqrt#1#2{%
\setbox0=\hbox{$#1\oldsqrt{#2\,}$}\dimen0=\ht0
\advance\dimen0-0.2\ht0
\setbox2=\hbox{\vrule height\ht0 depth -\dimen0}%
{\box0\lower0.4pt\box2}}

\begin{document}
\begin{titlepage}
\date{today}       \hfill
{\raggedleft LMU-ASC 58/15}
\begin{center}
\vskip .5in
{\Large \bf  Reflection and transmission of conformal perturbation defects}\\
 
\vskip .250in

\vskip .5in
{\large Ilka Brunner${^{1}}$ and Cornelius Schmidt-Colinet${}^{2}$ }

\vskip 0.5cm
{\it Arnold Sommerfeld Center for Theoretical Physics\\
Ludwig-Maximilians-Universit\"at M\"unchen\\
Theresienstra{\ss}e 37, 80333 Munich, Germany
}
\end{center}

\vskip .5in

\begin{abstract} \large
We consider reflection and transmission of interfaces which implement renormalisation group 
flows between conformal fixed points in two dimensions. Such an RG interface is constructed 
from the identity defect in the ultraviolet CFT by perturbing the theory on one side of the defect line. 
We compute reflection and transmission coefficients in perturbation theory to third order in the coupling 
constant and check our calculations against exact constructions of RG interfaces between coset models.
\end{abstract}
\vfill
{\small ${}^1$  {\tt ilka.brunner@physik.uni-muenchen.de}\\
${}^2$ {\tt schmidt.co@physik.uni-muenchen.de}}

\end{titlepage}
\section{Introduction}
Interfaces between two-dimensional conformal field theories 
\cite{Oshikawa:1996ww,Petkova:2000ip,Bachas:2001vj} play an important role in statistical mechanics 
and string theory. In some sense they generalise the notion of a conformal field theory in a system 
with boundary by allowing a transmission of energy and momentum \cite{QRW}, entropy 
\cite{Friedan:2005ca,Friedan:2005bz}, or other conserved quantities \cite{Kimura:2014hva} across 
the boundary into another critical system. 
From this point of view, transmission and reflection provide a fundamental reason for studying 
interfaces, with applications arising in fields such as the study of scattering properties through 
junctions or impurities in $1+1$ dimensional conformal systems (for recent work in this direction 
see {\it e.g.} \cite{Affleck2014,Bernard2015}).

Similarly as in the boundary case, there are conditions for field configurations on the interface 
which preserve some part of the symmetry algebra and render the interface conformal. The local 
condition for a conformal interface (or defect) is that the flow of energy or momentum parallel to 
the interface is continuous. If the interface separates the theories $CFT^{(1)}$ and 
$CFT^{(2)}$ along the real line, this condition reads
\begin{equation}\label{conformalitycondition}
T^{(1)}-\tilde{T}^{(1)}=T^{(2)}-\tilde{T}^{(2)}\,,
\end{equation}
where $T^{(i)}$ and $\tilde{T}^{(i)}$ are the holomorphic and antiholomorphic components
of the energy-momentum tensor in $CFT^{(i)}$ ($i=1,\,2$). The conformal boundary condition 
satisfies \eqref{conformalitycondition} by setting either side of the equation to zero. On the 
other hand, a solution equating the holomorphic components across the interface leads to a 
topological defect, which can be moved and deformed in the system at no cost of energy. An 
immediate consequence of \eqref{conformalitycondition} is that the difference between left- and 
right-moving Virasoro central charges matches for the two theories. All conformal interface 
conditions are conformal boundary conditions by the ``doubling'' or ``folding'' trick 
\cite{Wong:1994np,Oshikawa:1996ww,QRW}, which maps the system on one side of the interface onto 
the other side and considers a suitably defined product CFT on the space bounded by the original 
interface.

Compared to topological defects, general conformal interfaces are less well understood. This applies in 
particular to interfaces separating CFTs with different central charges. One interesting class of such interfaces 
implements relevant renormalisation group flows, where the CFT on one side of the interface admits a perturbation 
that flows in the infrared to the theory on the other side \cite{Brunner:2007ur}. Examples of such interfaces 
corresponding to relevant RG flows have been constructed in the context of $AdS_3/CFT_2$ in terms of Janus 
solutions, which interpolate between different embeddings of $su(2)$ into $su(N)$ in the Chern-Simons formulation 
\cite{Gutperle:2013ema}. In \cite{Gaiotto}, an exact construction was proposed for RG interfaces 
between Virasoro Minimal Models of adjacent levels, corresponding to the well-known flows studied in
\cite{Zamolodchikov:1987ti}. This construction was generalised in \cite{Poghosyan2}
to flows between general maximal-embedding coset models. In the case where the flow is 
perturbatively tractable, the RG interface can be obtained as a conformal perturbation defect 
\cite{Konechny:2014opa} by restricting the domain of the perturbation in the orginal UV theory.

As non-local linear operators, conformal interfaces encode relations between the adjacent theories. 
This includes the more intuitive case where symmetries and dualitites of a particular CFT are described 
by topological defects \cite{Petkova:2000ip,Frohlich:2006ch}, but also more vaguely a notion of how 
close two CFTs are to each other. As an example for the latter aspect we mention the idea that the space of all
two-dimensional CFTs may admit a distance measure based on the entropy of certain interfaces between any 
two theories \cite{BBDR}. In the context of RG interfaces it was pointed out in 
\cite{Gukov:2015qea} that their classification provides a concrete realisation of the counting of 
RG flows between fixed points.

In the study of RG interfaces we are interested in easily accessible indicators,
{\it i.e.} physical quantities that characterise the interface condition. Besides the interface 
entropy, another such quantity is provided by the reflection/transmission property.

In this paper we consider the reflection and transmission of energy as defined in \cite{QRW}
for the case of RG interfaces. Correlation functions of the energy-momentum tensor lead to a reflection 
coefficient ${\cal R}$ and a transmission coefficient ${\cal T}$, related by ${\cal R}+{\cal T}=1$.
In unitary theories ${\cal T}\in(0,1)$, and the coefficients have the intuitive property that the 
transmission is equal to 1 for the totally transmissive topological defects, and vanishes for totally 
reflective boundary conditions.

The definition of the reflection and transmission coefficients can be extended to the RG trajectory. 
Starting from a totally transmissive identity defect, we expect the transmissivity to decrease (resp. 
the reflectivity to increase) along the flow. After explaining our setup and notations in
section~\ref{section:setup}, we confirm this expectation perturbatively for all relevant and marginal 
flows in section~\ref{section:RT}. We compare our calculation at the fixed points with the perturbative 
formula for the entropy of the RG interface \cite{Konechny:2014opa}, and find that the reflection 
coefficient is related to first order in the simple way \eqref{gandR} to the entropy of the interface. 
In section~\ref{section:GP defects} we consider the RG defects between coset models mentioned 
above, and test our perturbative calculation against the exact coefficients obtained from these constructions. 
We briefly consider marginal deformations of the free boson
in section~\ref{section:limitcases}, and conclude in section~\ref{section:conclusion}.
Some technical steps of sections~\ref{section:RT} and \ref{section:GP defects} are collected in the appendix.


\section{Setup and definitions}\label{section:setup}
\begin{figure}[hb]
\begin{center}
\begin{pspicture}(0,0)(9,2.5)
\rput(4.8,0.5){\includegraphics[width=9cm]{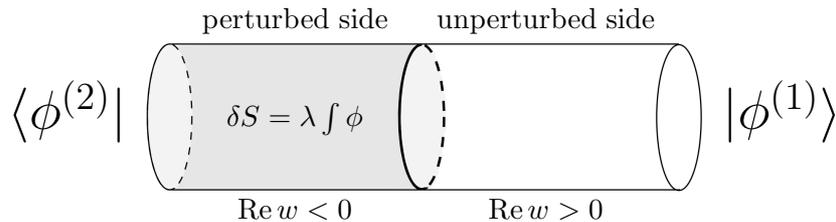}}
\rput(0,0.8){{\huge $\bra{\phi^{(2)}}$}}
\rput(3,0.7){$\delta S=\lambda\int\phi$}
\rput(9.4,0.8){{\huge $\ket{\phi^{(1)}}$}}
\rput(3,2){perturbed side}
\rput(6.3,2){unperturbed side}
\rput(3,-0.5){$\re w <0$}
\rput(6.3,-0.5){$\re w >0$}
\end{pspicture}
\end{center}
\caption{\it Our setup on a cylinder of circumference $\beta$. The defect wraps the cylinder at $\re w=0$.
Asymptotically far away from it, the system is in the states $\ket{\phi^{(1)}}$ of the UV
and $\ket{\phi^{(2)}}$ of the IR theory, respectively. In perturbation theory, $\phi^{(2)}$ will be given by a 
perturbed UV operator.}\label{Setup}
\end{figure}

For reasons of later convenience we consider the system on a torus, split into two cylindrical halves. 
On one half the system is at its UV fixed point, on the other half it is described by an IR CFT
obtained from the UV by a relevant but almost marginal perturbation. The torus is stretched such 
that we can describe the region around one of the interfaces in terms of a long cylinder of 
circumference~$\beta$. At the ends of the cylinder we will prescribe asymptotic states 
$\ket{\phi^{(1)}},\, \bra{\phi^{(2)}}$ of the UV and IR theory, respectively 
(see Figure~\ref{Setup}). We will use coordinates $w$ on the cylinder geometry.

We will assume that the IR fixed point is obtained from a perturbation of the UV fixed point by 
the scalar operator $\phi$ of conformal dimension $\Delta=2-\delta$, with $0\leq\delta\ll 2$. The UV 
action is perturbed by a term
\begin{equation}\label{cylinderperturbation}
\delta S = \lambda(\tfrac{2\pi}{\beta})^{\delta}\int d^2w\,\phi(w)\,.
\end{equation}
The explicit factor of $2\pi/\beta$ is part of our scheme choice \cite{Konechny:2014opa}. The 
coupling constant $\lambda$ is dimensionless. We will write $\phi(w)$ instead of $\phi(w,\bar{w})$ 
for simplicity. 

We will consider situations where the OPE of $\phi$ with itself is of the form
\begin{equation}\label{phiOPE}
\phi(w)\phi(0) = |w|^{-2\Delta} \,+\,C\,\phi(0)\,|w|^{-\Delta}\,+\,{\rm irrelevant}\,.
\end{equation}
This ensures that to the order of perturbation theory we will be interested in, no other couplings
enter the beta function of the perturbing operator $\phi$.\footnote{Allowing for other almost-marginal
operators that mix with $\phi$ is possible (for explicit examples see {\it e.g.} \cite{Crnkovic,Gaberdiel:2009hk}), 
but only renders our discussion unnecessarily tedious.}
If we use regularisation by a position-space cut-off, the beta function for the 
renormalised coupling constant reads
\begin{equation}\label{betafunction}
\beta=\delta\,\lambda \,+\, \pi\,C\,\lambda^2\,+\,\pi^2\,D\,\lambda^3\,+\,\mathcal{O}(\lambda^4)\,, 
\end{equation}
where $C$ is the OPE coefficient in \eqref{phiOPE}. For $\delta>0$ will assume that $C>0$, and 
in fact that $C$ and $D$ are of order~1, such that in particular $\delta/C\ll 1$. In this case the flow admits 
an IR fixed point perturbatively close to the original UV fixed point. In the IR the value of the 
renormalised coupling constant is of the order of the anomalous dimension $\delta$,
\begin{equation}\label{lambdaIR}
\lambda_{IR}=-\frac{\delta}{\pi\,C}-\frac{D\,\delta^2}{\pi\,C^3}+\mathcal{O}(\delta^3)\,,
\end{equation}
and expansion in $\lambda_{IR}$ (or, equivalently, $\delta$) is valid at the new
fixed point~\cite{Zamolodchikov:1987ti,Ludwig:1987gs}.
 
Recall that the value of correlation functions of a collection of (renormalised) local operators 
$\mathcal{O}$ in the perturbed theory is given by
\begin{equation}
\cor{\mathcal{O}}_{\rm pert}=\langle\mathcal{O}e^{\delta S}\rangle=\cor{\mathcal{O}}
+\cor{\mathcal{O}\,\delta S}+\cor{\mathcal{O}\,(\delta S)^2}+\ldots\,,
\end{equation}
where correlation functions without any subscripts denote those of the UV CFT. 

All our computations will effectively be performed on the plane with coordinates 
\mbox{$z=\exp(\tfrac{2\pi}{\beta}w)$}. In these coordinates, the interface is wrapped around
the unit circle. Unless otherwise stated, all correlation functions
in the following will be understood with respect to these planar coordinates.
The perturbation \eqref{cylinderperturbation} can be written on the plane as
\begin{equation}\label{planeperturbation}
\delta S = \lambda\int d^2z\,|z|^{-\delta} \phi(z)\,.
\end{equation}
Observe that this perturbation is invariant under $z\rightarrow1/z$.
In the presence of the defect, the integral only runs over $\re w<0$, {\it i.e.} the 
unit disc in the coordinates $z$. 

We will regularise UV divergences by a 
position-space cutoff $\epsilon_w\equiv\tfrac{\beta}{2\pi}\epsilon$ on the cylinder, with 
$0<\epsilon\ll 1$. Besides the UV cut-off we will also need a large-distance cut-off $L$ on the cylinder, 
such that $-\tfrac{\beta}{2\pi}L<\re w$. All cut-offs need to be transformed 
accordingly when we change coordinates. However, in all coordinates we are about to employ it will be 
sufficient to keep only the lowest order in the expansions of small $\epsilon$ (or large $L$), 
and thus use circular cut-offs.

\section{Perturbative calculation of reflection and transmission}\label{section:RT}
A definition for measuring reflection and transmission across a conformal interface was given 
in \cite{QRW} (see also \cite{Kimura:2015nka} for a recent refinement to multi-junctions). The 
definition is based on the correlation of energy-momentum tensor components on the plane. For a conformal 
interface wrapped around the unit circle, separating two fixed points $CFT^{(1)}$ and $CFT^{(2)}$, we define the unitary matrix
\begin{equation}\label{Rdefinition}
R\,=\,\frac{1}{\cor{0^{(2)}|0^{(1)}}}\left(\begin{array}{cc}
\cor{0^{(2)}|T^{(1)}\tilde{T}^{(1)}} & \cor{T^{(2)}|T^{(1)}} \\
\cor{\tilde{T}^{(2)}|\tilde{T}^{(1)}} & \cor{T^{(2)}\tilde{T}^{(2)}|0^{(1)}}
\end{array}\right)\equiv
\left(\begin{array}{cc}
R_{11} & R_{12} \\
R_{21} & R_{22}
\end{array}\right)
\,.
\end{equation}
Here $T^{(i)},\,\tilde{T}^{(i)}$ are the holomorphic and antiholomorphic 
energy-momentum tensor components of the plane, and $\ket{0^{(i)}}$ is the vacuum state in $CFT^{(i)}$.
Reflection ${\cal R}$ and transmission ${\cal T}$ are defined by means of the matrix 
$R$ as
\begin{equation}\label{RTdefinition}
{\cal R}= \mathcal{N}^{-1}(R_{11}+R_{22})\,,\quad
{\cal T}= \mathcal{N}^{-1}(R_{12}+R_{21})\,,
\end{equation}
where $\mathcal{N}=\sum_{i,j}R_{ij}$. Obviously we have the intuitive relation ${\cal R}+{\cal T} = 1\,.$
As was shown in \cite{QRW}, the matrix $R$ is actually fixed by conformal symmetry up to a single free 
parameter, determined by the precise interface condition. Here we will keep working with the 
more intuitve matrix $R$. As explained in \cite{QRW}, its entries are closely related 
to the transmission of (bulk) entropy through quantum wire junctions as considered in 
\cite{Friedan:2005ca,Friedan:2005bz}. \\

The definition of the matrix $R$ can be extended to RG trajectories between conformal fixed points. In 
the following we compute $R$ for a perturbative conformal defect on our cylinder geometry. After 
perturbation, the entries of $R$ will be of the form
\begin{equation}\label{Rperturbativedefinition}
R_{ij}=R_{ij}^{(0)}\,+\,\lambda^2\,R_{ij}^{(2)}\,+\,\lambda^3\,R_{ij}^{(3)}\,+\,
\mathcal{O}(\lambda^4)\,.
\end{equation}
Since we are interested mostly in the result close to the IR fixed point, we will further expand the 
coefficients of each order of $\lambda$ in the parameter $\delta$, and only keep the terms necessary 
for a consistent expansion in the IR. More concretely this means that $R_{ij}^{(2)}$ will be expanded 
to the first subleading and $R_{ij}^{(3)}$ to the leading order in $\delta$. Notice that at the fixed 
points the normalisation constant in \eqref{RTdefinition} is given by the disc one-point 
function of the bulk operator $T\tilde{T}$ in the folded picture,
\begin{equation}\label{normalisationfactor}
\mathcal{N}=\sum_{i,j}R_{ij}=\frac{\langle T\tilde{T}(0)\rangle_{\rm disc}}%
{\cor{1}_{\rm disc}}=(c^{(1)}+c^{(2)})/2\,.
\end{equation}
Here $c^{(i)}$ denotes the central charge of $CFT^{(i)}$. Under perturbation, the normalisation 
constant ${\cal N}$ is therefore determined from the change in the central charge of the theory on 
the perturbed side of the defect. The perturbative change of the central charge was computed in 
\cite{Ludwig:1987gs}, in the same scheme as we are employing here. It is therefore possible to derive 
the value of $R_{12}=R_{21}$ from the values of $R_{11}$ and $R_{22}$ not only at the fixed points, 
but for the full perturbative result in our scheme. We will calculate the perturbative change of 
$R_{11}$ and $R_{22}$ in the sections~\ref{section:secondorder} and \ref{section:thirdorder}, and 
turn to the off-diagonal entries after that in section~\ref{section:Perturbative Result}, where we 
also collect the results of the perturbative calculation.

Notice that $R_{11}^{(0)}=R_{22}^{(0)}=0$, such that the
perturbative change of the one-point function in the numerator up to the order $\lambda^3$
drops out in the calculation of the diagonal entries.

\subsection{Second order}\label{section:secondorder}
The coefficient of $\lambda^2$ in $R_{11}$ reads
\begin{equation}
R_{11}^{(2)}=\frac{1}{2}\int d^2z_1d^2z_2|z_1z_2|^{-\delta}\cor{T(\infty)\tilde{T}(\infty)
\phi(z_1)\phi(z_2)}\,,
\end{equation}
with
\begin{equation}
\cor{T(\infty)\tilde{T}(\infty)\phi(z_1)\phi(z_2)}=\tfrac{\Delta^2}{4}|z_{21}|^{2\delta}\,.
\end{equation}
We choose $|z_2|\leq|z_1|$ at the cost of an additional factor of 2,
and set $z_2=\xi z_1$ for $|\xi|\leq 1$. Then we have
\begin{equation}
R_{11}^{(2)}=\frac{\Delta^2}{4}\int d^2z_1\,|z_1|^2\int d^2\xi\,|\xi|^{-\delta}\,|1-\xi|^{2\delta}\,.
\end{equation}
We expand
\begin{equation}\label{square-term expansion}
|1-\xi|^{2\delta}=1+2\delta\log|1-\xi|+\mathcal{O}(\delta^2)
\end{equation}
and notice that the term proportional to $\log|1-\xi|$ drops out by the angular integration.
Without the need of a cutoff in this particular calculation we find 
\begin{align}\label{R11order2result}
R_{11}^{(2)}
=\frac{\pi^2}{2}-\frac{\pi^2}{4}\delta+\mathcal{O}(\delta^2)\,.
\end{align}
The second-order coefficient in $R_{22}$ reads
\begin{equation}\label{secondordercoef}
R_{22}^{(2)}=\frac{1}{2}\int d^2z_1d^2z_2\,|z_1z_2|^{-\delta}\cor{\phi(z_1)\phi(z_2)T(0)\tilde{T}(0)}\,,
\end{equation}
with
\begin{equation}\label{R222cor}
\cor{\phi(z_1)\phi(z_2)T(0)\tilde{T}(0)}=\frac{\Delta^2}{4}\frac{|z_{21}|^{2\delta}}{|z_1z_2|^4}\,.
\end{equation}
There are now power-law divergences when the $\phi$ insertions at $z_1$ and $z_2$
approach the energy-momentum tensor at the origin, while
the situation where the $\phi$ insertions are close to 
each other is still suppressed. The coordinate 
transformation $z_2=\xi z_1$ gives
\begin{equation}
\int d^2z_1d^2z_2\,\frac{|z_{21}|^{2\delta}}{|z_1z_2|^{4+\delta}}=
2\int d^2z_1|z_1|^{-6}\int d^2\xi|\xi|^{-4-\delta}|1-\xi|^{2\delta}\,.
\end{equation}
Using \eqref{square-term expansion} again one finds
\begin{align}
\int d^2z_1d^2z_2\,\frac{|z_{21}|^{2\delta}}{|z_1z_2|^{4+\delta}}&=
2\int d^2z_1|z_1|^{-6}\int d^2\xi|\xi|^{-4-\delta}\,+\,\mathcal{O}(\delta^2)\,.
\end{align}
Notice that the cut-off for $\xi\rightarrow 0$ is 
$\epsilon+\mathcal{O}(\epsilon^2)$. Using minimal subtraction we obtain
\begin{align}
\int d^2z_1d^2z_2\,\frac{|z_{21}|^{2\delta}}{|z_1z_2|^{4+\delta}}&=
-\frac{4\pi}{2+\delta}\int d^2z_1|z_1|^{-6}\left(1-\epsilon^{-2-\delta}\,+\,\mathcal{O}
(\delta^2)\right)\nonumber\\
&=\pi^2-\frac{\pi^2}{2}\delta+\mathcal{O}(\delta)\,.
\end{align}
Restoring the prefactors from \eqref{secondordercoef} and \eqref{R222cor} we therefore have
\begin{equation}\label{R22order2result}
R_{22}^{(2)}=\frac{\pi^2}{2}\,-\,\frac{3\pi^2}{4}\delta\,+
\,\mathcal{O}(\delta^2)\,.
\end{equation}

\subsection{Third order}\label{section:thirdorder}
Recall that in order to calculate quantities at the IR fixed point to third order in the value of the 
coupling constant, we only need to compute the leading order contributions in $\delta$ of the third-
order coefficients $R_{ij}^{(3)}$. The third-order coefficient in $R_{11}$ is given by
\begin{equation}
R_{11}^{(3)}=\frac{1}{6}\int d^2z_1d^2z_2d^2z_3|z_1z_2z_3|^{-\delta}\cor{T(\infty)\tilde{T}
(\infty)\phi(z_1)\phi(z_2)\phi(z_3)}\,,
\end{equation}
where
\begin{align}
&\cor{T(\infty)\tilde{T}(\infty)\phi(z_1)\phi(z_2)\phi(z_3)}=
\frac{\Delta^2}{4}|z_1^2+z_2^2+z_3^2-z_1z_2-z_2z_3-z_3z_1|^2\cor{\phi(z_1)\phi(z_2)\phi(z_3)}\,,
\nonumber\\
&{\rm with}\qquad\cor{\phi(z_1)\phi(z_2)\phi(z_3)}=C|z_{12}z_{23}z_{31}|^{-\Delta}\,.
\end{align}
To leading order in $\delta$ the expression for $R_{11}^{(3)}$ simplifies to
\begin{align}\label{R11order3}
R_{11}^{(3)}&=\frac{\Delta^2}{24}C\int d^2z_1d^2z_2d^2z_3
\left|\frac{1}{z_{12}}+\frac{1}{z_{23}}+\frac{1}{z_{31}}\right|^2\nonumber\\
&=\frac{\Delta^2}{24}C\int d^2z_1d^2z_2d^2z_3
\left(3\left|\frac{1}{z_{12}}\right|^2+\frac{6}{z_{12}\bar{z}_{23}}\right)\,.
\end{align}
We defer the details of performing this integral to appendix~\ref{appendix:perturbativecalculations},
and will only state the result here. As might be intuitively clear from \eqref{R11order3}, the 
integration of the first summand in the bracket of the last line is a pure counterterm in our scheme,
\begin{equation}\label{R11order3firstpart}
\int d^2z_1d^2z_2d^2z_3\frac{3}{|z_{21}|^2}=-6\pi^3\log\epsilon\,+\,\mathcal{O}(\epsilon)\,.
\end{equation}
For the computation of the other contribution to \eqref{R11order3}, no cut-off is necessary,
and one obtains 
\begin{equation}\label{R11order3secondpart}
\int d^2z_1d^2z_2d^2z_3\frac{1}{z_{12}\bar{z}_{23}}=-\frac{\pi^3}{2}\,.
\end{equation}
Combining \eqref{R11order3firstpart} and \eqref{R11order3secondpart}, 
$R_{11}^{(3)}$ is therefore, to leading order in $\delta$,
\begin{equation}\label{R11order3result}
R_{11}^{(3)}=\frac{\Delta^2}{24}C\int d^2z_1d^2z_2d^2z_3
\left(\frac{3}{|z_{12}|^2}+\frac{6}{z_{12}\bar{z}_{23}}\right)
=-C\pi^3\log\epsilon\,-\,\tfrac{1}{2}C\pi^3\,+\,\mathcal{O}(\epsilon)\,.
\end{equation}
Finally, the coefficient of the third-order contribution to $R_{22}$ reads
\begin{equation}
R_{22}^{(3)}=\frac{1}{6}\int d^2z_1d^2z_2d^2z_3|z_1z_2z_3|^{-\delta}
\cor{\phi(z_1)\phi(z_2)\phi(z_3)T(0)\tilde{T}(0)}\,.
\end{equation}
Here,
\begin{equation}
\cor{\phi(z_1)\phi(z_2)\phi(z_3)T(0)\tilde{T}(0)}=\frac{\Delta^2\,C}{4}\,
\frac{|z_1^2z_2^2+z_2^2z_3^2+z_3^2z_1^2-z_1z_2z_3(z_1+z_2+z_3)|^2}{|z_1\,z_2\,z_3|^4\,%
|z_{12}z_{23}z_{31}|^\Delta}\,.
\end{equation}
We are again only interested in the leading-order term in $\delta$.
Similarly as in the computation of $R_{11}$ we order $|z_1|\leq|z_2|\leq|z_3|$
at the cost of an additional factor of 6, and replace
\begin{equation}
z_2=\xi z_3\,,\qquad z_1=\xi\eta z_3\,.
\end{equation} 
The integral one has to calculate then becomes
\begin{align}\label{longint}
&\int d^2z_1d^2z_2d^2z_3\left|\frac{z_1^2z_2^2+z_2^2z_3^2+z_3^2z_1^2-z_1z_2z_3(z_1+z_2+z_3)}
{z_1^2\,z_2^2\,z_3^2\,%
z_{12}z_{23}z_{31}}\right|^2\nonumber\\
&\quad= 6\int d^2 z_3d^2\xi d^2\eta|z_3|^{-6}|\xi\eta|^{-4}\left|
\frac{1}{1-\xi}+\frac{\eta}{1-\eta}-\frac{\eta}{1-\eta\xi}\right|^2\\
&\quad=6\int d^2 z_3d^2\xi d^2\eta|z_3|^{-6}|\xi\eta|^{-4}\bigg(
\frac{1}{|1-\xi|^2}+\frac{|\eta|^2}{|1-\eta|^2}+\frac{|\eta|^2}{|1-\xi\eta|^2}\nonumber\\
&\qquad\qquad+\frac{2\bar{\eta}}{(1-\xi)(1-\bar{\eta})}-\frac{2\bar{\eta}}{(1-\xi)
(1-\bar{\xi}\bar{\eta})}
-\frac{2|\eta|^2}{(1-\eta)(1-\bar{\xi}\bar{\eta})}\bigg)\,.\nonumber
\end{align}
The integration can be performed with the same methods as before for $R_{11}$; the interested reader 
can find more details in appendix~\ref{appendix:perturbativecalculations}. 
Eventually the result of the integration of \eqref{longint} yields
\begin{equation}\label{R22order3result}
R_{22}^{(3)}=-\frac{\pi^3}{2}\, C\,.
\end{equation}

\subsection{Perturbative Result}\label{section:Perturbative Result}
The perturbative result for $R_{11}$ and $R_{22}$ up to third order in the coupling constant $\lambda$ 
is now obtained from \eqref{Rperturbativedefinition}, \eqref{R11order2result}, 
\eqref{R22order2result}, \eqref{R11order3result}, and \eqref{R22order3result}:
\begin{align}\label{pertresult part1}
R_{11}&=\left(\frac{\pi^2}{2}-\frac{\pi^2}{4}\delta\right)\lambda^2\,-\,
\frac{\pi^3}{2}C\lambda^3\,,\nonumber\\
R_{22}&=\left(\frac{\pi^2}{2}-\frac{3\pi^2}{4}\delta\right)\lambda^2\,-\,\frac{\pi^3}{2}C\lambda^3\,.
\end{align}
As mentioned before, the coefficients are expanded such that the final result at the fixed point is 
valid up to third order in the anomalous dimension $\delta$. We observe that $R_{22}$, which measures the 
reflection on the IR side of the RG defect, is smaller than the reflection measured by $R_{11}$ on the UV 
side. This illustrates the notion of information loss along the RG flow --- intuitively, modes sent towards 
the defect from the UV side have a lower chance of finding a suitable form for transmission than vice 
versa. Notice also that for the case of marginal flows ($\delta=0$), the expressions for $R_{11}$ and 
$R_{22}$ coincide.

As mentioned in the beginning of section~\ref{section:RT}, we can compute the perturbative 
result for $R_{12}=R_{21}$ from the change in the central charge. Employing the same scheme as we do 
here, the general result for the perturbed central charge was obtained in \cite{Ludwig:1987gs},
\begin{equation}
c^{\rm pert}=c\,-\,3\pi^2\delta\lambda^2\,-\,2\pi^3 C\lambda^3\,+\,\mathcal{O}(\lambda^4)\,.
\end{equation}
From \eqref{normalisationfactor} we see that
\begin{equation}
R_{12}=R_{21}=\tfrac{1}{4}(c^{(1)}+c^{(1)\,{\rm pert}})\,-\,\tfrac{1}{2}\,(R_{11}+R_{22})\,,
\end{equation}
such that up to third order in $\lambda$ we have
\begin{equation}\label{pertresult part2}
R_{12}=R_{21}=\frac{c^{(1)}}{2}\,-\,\left(\frac{\pi^2}{2}+\frac{\pi^2}{4}\,\delta\right)%
\lambda^2\,.
\end{equation}
Notice that the coefficient of $\lambda^3$ vanishes to leading order in $\delta$, which means that
to this order the coefficient is given by a pure counterterm in our scheme. We can now combine 
\eqref{pertresult part1} and \eqref{pertresult part2} to compute the reflection and transmission 
coefficients from \eqref{RTdefinition}:
\begin{equation}\label{pertresult part3}
{\cal R}\,=\,\left(\frac{\pi^2}{c^{(1)}}-\frac{\pi^2}{c^{(1)}}\,\delta\right)\lambda^2\,-\,
\frac{\pi^3 C}{c^{(1)}}\,\lambda^3\,,\quad
{\cal T}\,=\,1\,-\,\left(\frac{\pi^2}{c^{(1)}}-\frac{\pi^2}{c^{(1)}}\,\delta\right)\lambda^2\,+\,
\frac{\pi^3 C}{c^{(1)}}\,\lambda^3\,.
\end{equation}
Let us write out the result for ${\cal R}$ at the fixed point. From the value of the coupling 
constant \eqref{lambdaIR} we obtain
\begin{equation}\label{pertresultR}
{\cal R}= \frac{2}{c^{(1)}}\left(\frac{\delta^2}{2C^2}+\frac{D\,\delta^3}{C^4}+\mathcal{O}(\delta^4)\right)\,.
\end{equation}
It turns out that to this order in perturbation, the reflection coefficient is related in a rather 
simple way to the entropy of the RG interface. The entropy is the logarithm of the $g$ factor 
\cite{Affleck:1991tk}, which corresponds to the overlap of the vacua across the interface. 
For RG interfaces, the perturbative calculation of the $g$ factor was done in 
\cite{Konechny:2014opa}, with the result
\begin{equation}\label{gandR}
g^2=1+\frac{\delta^2}{2C^2}+\frac{\delta^3 D}{C^4}+\mathcal{O}(\delta^4) = 
1+\frac{c^{(1)}}{2}\,{\cal R}+\mathcal{O}(\delta^4)
\end{equation}
at the IR fixed point.
To lowest order, the transmission and the interface entropy thus contain the same information.
We emphasise however that the entropy cannot be a universal function of the transmission
to higher orders. One intuitive reason is that the reflection is determined only from the 
vacuum representation, while the $g$ factor is involved in Cardy's condition, and thus
in general must contain more subtle information about the CFT. We will see an explicit 
example for this in the next section.

\section{Gaiotto-Poghosyan defects}\label{section:GP defects}
In this section we check our perturbative result \eqref{pertresult part1}, \eqref{pertresult part2},
\eqref{pertresult part3}, and \eqref{gandR} for flows between CFTs whose chiral algebra is a 
maximally-embedded coset of the affine algebra $\hat{a}$ associated to the simple Lie algebra $a$. 
The coset algebra reads 
\begin{equation}\label{cosetalgebra}
M_{k,l}=\frac{\hat{a}_k\oplus \hat{a}_l}{\hat{a}_{k+l}}\,.
\end{equation}
We consider the diagonal modular invariant. The constituent WZW model based 
on the affine algebra $\hat{a}_k$ has central charge
\begin{equation}
c_{k}=\frac{{\rm dim}(a)\,k}{k+\dcox_a}\,,
\end{equation}
where $\dcox$ denotes the dual Coxeter number of $a$. The coset CFT 
\eqref{cosetalgebra} therefore has central charge
\begin{equation}
c_{k,l}=\frac{{\rm dim}(a)\,l}{l+\dcox_a}\left(
1-\frac{\dcox_a(l+\dcox_a)}{(k+\dcox_a)(k+l+\dcox_a)}\right)\,.
\end{equation}
For $k>l$ these coset CFTs admit perturbations leading to massless theories, with RG flows between the 
fixed points \cite{Ahn:1990gn}
\begin{equation}\label{hoppingflow}
M_{k,l}\rightarrow M_{k-l,l}\,.
\end{equation}
A well-known instance of such a sequence of flows exists between the Virasoro Minimal 
Models, where $a=su(2)$, $l=1$ \cite{Zamolodchikov:1987ti}.
The perturbing field in the UV theory is the primary coset operator in the representation 
$(0,0;{\rm adj})$. It has conformal dimension
\begin{equation}\label{DeltaUV}
\Delta=2-\frac{2\dcox}{k+l+\dcox}\,.
\end{equation}
Along the flow, the perturbation does not mix with other relevant fields to all orders in 
perturbation theory. From the IR point of view, in cases where $k>l+1$ the flow is described 
by an irrelevant perturbation by the operator associated with the representation $({\rm adj},0;0)$
of $M_{k-l,l}$. This operator has the conformal dimension
\begin{equation}\label{DeltaIR}
\Delta_{IR}=2+\frac{2\dcox}{k-l+\dcox}\,.
\end{equation}
The RG interfaces for these flows have been worked out in \cite{Gaiotto} for the case of the Virasoro 
Minimal Models, and generalised to the flow \eqref{hoppingflow} in \cite{Poghosyan2}. Following
\cite{Gaiotto}, we briefly repeat the construction in the folded picture, where the defect corresponds
to a boundary state on the unit circle. From basic properties of topological defects under the RG 
flow \eqref{hoppingflow}, one deduces that the defect must be in a class which preserves specific
symmetry of the folded theory $M_{k-l,l}\oplus M_{k,l}$. As a first result it was noted in 
\cite{Gaiotto} that the only
non-trivial one-point functions in the doubled theory with the boundary condition corresponding
to the RG defect correspond to a projected sector. The projection $P$ is onto states which have the 
same representation label in the two copies of the algebra $\hat{a}_k$ --- one appearing in the 
numerator of $M_{k,l}$, and the other appearing in the denominator of $M_{k-l,l}$. 
The projection is thus given by a map
\begin{equation}\label{superalgebra}
M_{k,l}\oplus M_{k-l,l}\,\rightarrow\, 
P(M_{k,l}\oplus M_{k-l,l}) \cong \frac{\hat{a}_{k-l}\oplus \hat{a}_l\oplus\hat{a}_l}{\hat{a}_{k+l}}\,.
\end{equation}
Let $r^{(n)}$ denote a representation of the affine chiral algebra $\hat{a}_n$, and we write
$(r^{(k)},r^{(l)};r^{(k+l)})$ for a representation of $M_{k,l}$ as before. A representation
$(r^{(k-l)},r^{(l)},s^{(l)};r^{(k+l)})$ of the right-hand side in \eqref{superalgebra} is the 
image of the direct sum\footnote{Selection rules are implicit in \eqref{Ponrepresentations}.}
\begin{equation}\label{Ponrepresentations}
(r^{(k-l)},r^{(l)},s^{(l)};r^{(k+l)})=
\bigoplus_{r^{(k)}}(r^{(k)},r^{(l)};r^{(k+l)})\otimes(r^{(k-l)},s^{(l)};r^{(k)})\,.
\end{equation}
The RG interface corresponds to a fusion product of a boundary condition for the chiral algebra 
on the right-hand side of \eqref{superalgebra} with a topological defect interpolating between the two 
sides.\footnote{In \cite{Gaiotto}, the algebra on the right-hand side of \eqref{superalgebra} was in fact
interpreted as a product of a supersymmetric Minimal Model and the Ising model. This 
interpretation follows a general pattern established in \cite{Crnkovic}.}
Let us first consider the boundary condition. It was argued in \cite{Gaiotto} that the boundary 
condition must preserve the symmetry in the way
\begin{equation}
\frac{\hat{a}_{k-l}}{\hat{a}_{k+l}}\oplus(\hat{a}_l\oplus \hat{a}_l)\,,
\end{equation}
{\it i.e.} it must correspond to a standard (Cardy) state on the coset part, multiplied with a 
$\mathbb{Z}_2$ permutation brane for the $\hat{a}_l$ factors. Such a state
has the form \cite{Recknagel:2002qq,Ishikawa:2001zu,Quella:2002ct}
\begin{equation}\label{GPboundarystate}
\braneket{R^{(k-l)},R^{(l)},R^{(l)},R^{(k+l)}}_{\mathbb{Z}_2}=\!\!\!
\sum_{\mbox{\tiny $\substack{r^{(k-l)},\\r^{(k+l)}}$}}
\frac{S^{(k-l)}_{R,r}\bar{S}^{(k+l)}_{R,r}}{\sqrt{S^{(k-l)}_{0,r}\bar{S}^{(k+l)}_{0,r}}}
\sum_{r^{(l)}}\frac{S^{(l)}_{R,r}}{S^{(l)}_{0,r}}
\ishiket{r^{(k-l)},r^{(l)},r^{(l)},r^{(k+l)}}_{\mathbb{Z}_2}\,.
\end{equation}
Here, $S^{(n)}_{r,s}$ is a modular S matrix element of the WZW model based on $\hat{a}_n$, where we dropped 
the superscripts on the representation indices for brevity. In the permutation part, the two representations 
of $\hat{a}_l$ are exchanged, forcing the respective representation labels to be equal. In our notation for 
the Ishibashi state and the boundary state we kept the information on the permutation part by the subscript 
$\mathbb{Z}_2$.

As mentioned before, the boundary state \eqref{GPboundarystate} must be fused with a topological defect 
interpolating between the left- and the right-hand side of \eqref{superalgebra}. The important consistency 
condition for such a topological defect is that the fusion product of the defect with its conjugate must 
be a linear superposition of standard Cardy defects with non-negative integer coefficients. As was demonstrated 
in~\cite{Gaiotto}, one solution to this constraint is an operator of the form
\begin{equation}\label{GPtopologicaldefect}
{\cal D}=\sum_{a,b}\sqrt{\frac{S_{b0}^{(\mathcal{B})}}{S_{a0}^{(\mathcal{A})}}}\,\|a \big| b\|\,,
\end{equation}
where we use $\mathcal{A}$ ($\mathcal{B}$) to refer to the left(right)-hand side of 
\eqref{superalgebra}, and $a$ ($b$) denote irreducible representations in $\mathcal{A}$ 
($\mathcal{B}$). The symbol $\|a|b\|$ stands for the Ishibashi operator which maps the representation $a$ 
to the copy of $a$ within the representation $b$.

From the perturbative results in the large-$k$ limit it was then conjectured in \cite{Gaiotto} that 
the RG defect corresponds to the boundary state
\begin{equation}\label{actualGPdefect}
\braneket{RG}={\cal D}\braneket{0^{(k-l)},0^{(l)},0^{(l)},0^{(k+l)}}_{\mathbb{Z}_2}\,.
\end{equation}
The construction specifies the overlap between an operator $\Phi^{(1)}$ in the UV and an operator 
$\Phi^{(2)}$ in the IR. Writing both operators in terms of their chiral components 
$\Phi=\phi\tilde{\phi}$, their overlap is given by a disc one-point function of the 
operator\footnote{In the constructions \cite{Gaiotto,Poghosyan2}, the position of the chiral 
components $\phi^{(2)}$ and $\tilde{\phi}^{(2)}$ is reversed. We believe that the construction here is 
more intuitive given that we have performed the folding trick. Since the original theories are assumed 
left-right symmetric, this has no effect on the formulae.}
\begin{equation}
\mathcal{O}=(\phi^{(1)}\tilde{\phi}^{(2)})(\tilde{\phi}^{(1)}\phi^{(2)})\,,
\end{equation}
interpreted in the theory with chiral algebra \eqref{superalgebra},
\begin{equation}\label{GPoverlap}
\cor{\Phi^{(1)}|\Phi^{(2)}} = \cor{\mathcal{O}}_B=
\mathcal{S}\cor{\mathbb{Z}_{2}\left(\phi^{(1)}\tilde{\phi}^{(2)}\right)
(\tilde{\phi}^{(1)}\phi^{(2)})}\,.
\end{equation}
In the last expression, we find from \eqref{GPboundarystate}, \eqref{GPtopologicaldefect} 
and~\eqref{actualGPdefect} that $\mathcal{S}=\sqrt{S^{(k-l)}_{0,r}S^{(k+l)}_{0,r}}/S_{0,r}^{(k)}$.\\

Notice that in hindsight, the symmetries that must be respected by the defect may not come
as a surprise. As was shown in \cite{Ahn:1990gn}, the perturbed theories generically still 
contain two (non-local) chiral symmetry currents, which commute with the perturbation and among 
themselves to all orders. One current is associated with the representation $({\rm adj},0;0)$ 
and thus with the algebra factor $\hat{a}_k$. The other one is associated with $(0,{\rm adj};0)$,
resp. the algebra factor $\hat{a}_l$. These currents generate the fractional 
supersymmetries in the perturbed quantum field theories \cite{Bernard:1990cw,Ahn:1990gn}. 
The coset space of states can be decomposed in terms 
of representations generated by these currents. The projection \eqref{superalgebra} 
corresponds to the preservation of the $\hat{a}_k$ current, and the $\mathbb{Z}_2$ condition to the 
preservation of the $\hat{a}_l$ current.\\

For the reflection and transmission coefficients we are interested in the overlap between the energy momentum 
tensor components. Denoting the generating currents of the algebra $\hat{a}_n$ by $J^{(n)}$, and suppressing 
algebra labels for the moment, the Sugawara construction yields
\begin{align}\label{Tdecomposition}
T^{UV}&=\frac{l\,J^{(k-l)}J^{(k-l)}}{2(k+\dcox)(k+l+\dcox)}+
\frac{l\,J^{(l)}J^{(l)}}{2(k+\dcox)(k+l+\dcox)}
+\frac{k\,\hat{J}^{(l)}\hat{J}^{(l)}}{2(k+\dcox)(k+l+\dcox)}\nonumber\\
&\qquad\quad+\frac{l\,J^{(k-l)}J^{(l)}}{(k+\dcox)(k+l+\dcox)}-
\frac{J^{(k-l)}\hat{J}^{(l)}}{k+l+\dcox}-\frac{J^{(l)}\hat{J}^{(l)}}{k+l+\dcox}\,,\\[6pt]
T^{IR}&=\frac{l\,J^{(k-l)}J^{(k-l)}}{2(k+\dcox)(k-l+\dcox)}+
\frac{(k-l)J^{(l)}J^{(l)}}{2(k+\dcox)(l+\dcox)}-\frac{J^{(k-l)}J^{(l)}}{k+\dcox}\,,\nonumber
\end{align}
where we have decorated the generators of the $\hat{a}_l$ factor in $M_{k-l,l}$ with a hat in order to 
distinguish them from the generators of the $\hat{a}_l$ factor in $M_{k,l}$. Analogous formulae with right-
moving currents $\tilde{J}$ hold for the antiholomorphic components $\tilde{T}^{UV}$, $\tilde{T}^{IR}$. 
Since the prefactor $\mathcal{S}$ in \eqref{GPoverlap} only depends on modular $S$ matrix elements and
therefore only on the conformal representations of the IR and the UV field, it drops out in the
computation of the $R$ matrix elements \eqref{Rdefinition}. The entries of the matrix $R$ are therefore
simply given by correlators of the algebra currents in the decomposition of the energy-momentum tensor 
components \eqref{Tdecomposition}. Keeping in mind that the $\mathbb{Z}_2$ action has the effect 
of interchanging $J^{(l)}$ and $\hat{J}^{(l)}$, we find
\begin{align}\label{CosetR}
R_{11}=\cor{\mathbb{Z}_2(T^{UV})\,T^{UV}}&\,=\,\frac{{\rm dim}(a)\,l^2\,(k+l+2\dcox)}%
{2(k+\dcox)^2(k+l+\dcox)}=
\frac{{\rm dim}(a)l^2}{2k^2}\,\left(1-\frac{\dcox}{k}+\mathcal{O}(k^{-2})\right)\,,\nonumber\\[6pt]
R_{12}=\cor{\mathbb{Z}_2(T^{UV})\,T^{IR}}&\,=\,\frac{{\rm dim}(a)\,l\,(k-l)\,(k+l+2\dcox)}%
{2(k+\dcox)^2(l+\dcox)}\nonumber\\
&\quad\,=\,\frac{{\rm dim}(a)l}{2(l+\dcox)}\,-\frac{{\dim}(a) l(l+\dcox)}{2k^2}\left(%
1-\frac{2\dcox}{k}+\mathcal{O}(k^{-2})\right)\,,\\[6pt]
R_{21}=\cor{\mathbb{Z}_2(T^{IR})\,T^{UV}}&\,=\,R_{12}\,,\nonumber\\[6pt]
R_{22}=\cor{\mathbb{Z}_2(T^{IR})\,T^{IR}}&\,=\,\frac{{\rm dim}(a)\,l^2\,(k-l)}%
{2(k+\dcox)^2(k-l+\dcox)}=
\frac{{\rm dim}(a)l^2}{2k^2}\,\left(1-\frac{3\dcox}{k}+\mathcal{O}(k^{-2})\right)\,.\nonumber
\end{align}
For these results we used in particular that 
$\cor{J^{(k)}J^{(k)}}\equiv\sum_{b}\langle J^{(k)b}J^{(k)}_b\rangle=k\,{\rm dim}(a)$.
Notice that for $k>l$ the entry $R_{11}$ is larger than $R_{22}$ in all cases, just as
in the perturbative result \eqref{pertresult part1}.

We now want to compare these expressions with our perturbative result \eqref{pertresult part1}, 
\eqref{pertresult part2}. In order to do this, we need to work out the coefficients of the beta 
function \eqref{betafunction} up to the required order $\lambda^3$. As mentioned in the beginning of 
section~\ref{section:setup}, in the cut-off scheme $C$ coincides with the OPE coefficient of the perturbing 
field. This coefficient can be computed exactly by the methods of \cite{Dotsenko:1984nm,Crnkovic}. 
However, for our perturbative purposes we already have just enough data to determine $C$ up to the required order 
in a simpler way. 
Consider a coset representative of the perturbing field in the numerator of $M_{k,l}$. 
The chiral fields
\begin{equation}
\varphi^a:=kJ^{(l)a}-lJ^{(k)a}
\end{equation}
are Virasoro-primary fields in the vacuum representation of the numerator CFT of the coset 
$M_{k,l}$, and transform as primary fields in the adjoint representation of the denominator. Thus 
each of them contains the chiral half of the $(0,0;{\rm adj})$ coset field. A full canonically 
normalised representative is then given by\footnote{In \eqref{numerator-field ansatz} summation 
over the indices is understood, and we raise and lower indices with the Killing form of the 
algebra $a$.}
\begin{equation}\label{numerator-field ansatz}
\varphi= \frac{1}{\sqrt{{\rm dim} a}kl(k+l)}\,\varphi^a\tilde{\varphi}_a\,.
\end{equation}
Its three-point function coefficient is easily determined to be
\begin{equation}
C_{\varphi\varphi\varphi}=\frac{2\dcox}{\sqrt{{\rm dim}\,a}}\frac{(l-k)^2}{kl(k+l)}\,.
\end{equation}
In the limit $k\rightarrow\infty$, the coset field $\phi$ asymptotically becomes a field in the
untwisted sector of the continuous orbifold, coinciding with the current-current 
deformation
\begin{equation}
\phi\rightarrow \frac{1}{l\sqrt{{\rm dim} a}}\,J^{(l)a}\tilde{J}^{(l)}_a\qquad (k\rightarrow\infty)
\end{equation}
in the WZW model based on $\hat{a}_l$. This is just the same limit as for our field $\varphi$ in 
\eqref{numerator-field ansatz}. Under the assumption that the limit behaves well on the level of 
fields \cite{Roggenkamp:2003qp}, the factor in $\varphi$ representing the denominator part
will become trivial in the limit $k\rightarrow\infty$. The three-point function of $\varphi$
splits into the three-point funtion of the coset part $\phi$ and the denominator part $\phi^{(k+l)}$ 
-- schematically,
\begin{equation}
\cor{\varphi\varphi\varphi}=\cor{\phi\phi\phi}\langle\phi^{(k+l)}\phi^{(k+l)}\phi^{(k+l)}\rangle\,.
\end{equation}
We now write \mbox{$\phi^{(k+l)}=1+\tfrac{1}{k}\phi^{(k+l)}_1$}, such that to first order in $1/k$,
\begin{equation}
\langle\phi^{(k+l)}\phi^{(k+l)}\phi^{(k+l)}\rangle=1+\tfrac{3}{k}\langle\phi^{(k+l)}_1\rangle=
1+\mathcal{O}(k^{-2})\,.
\end{equation}
Therefore we have at least\footnote{In the case of the Virasoro Minimal models for example,
$C_{\varphi\varphi\varphi}$ is in fact correct up to terms of order $1/k^4$.}
\begin{equation}\label{cosetOPEcoefficientC}
C=C_{\varphi\varphi\varphi}+\mathcal{O}(k^{-2})=\frac{2\dcox}{\sqrt{{\rm dim}\,a}}
\left(\frac{1}{l}-\frac{3}{k}\right)+\mathcal{O}(k^{-2})\,.
\end{equation}
We can also determine the coefficients in the $\beta$ function from the dimension \eqref{DeltaIR} 
of the perturbing field in the IR. This dimension is given by the derivative of the beta function at the IR 
fixed point,
\begin{equation}
\Delta_{IR}=2-\partial_\lambda\beta\,\big|_{\lambda=\lambda_{IR}}\,.
\end{equation}
We use this to derive the expression for $D$, for which we only need the leading order in $1/k$. Inserting
\eqref{lambdaIR} for the value of the coupling at the IR fixed point we obtain
\begin{equation}\label{cosetcoefficientD}
D\,=\,-\frac{C^2(\Delta_{IR}-2)}{\delta^2}+\frac{C^2}{\delta}+\mathcal{O}(\delta)\,=\,
-\frac{l}{\dcox}\,C^2 \,+\,\mathcal{O}(k^{-1})\,=\,
-\frac{4\dcox}{{\rm dim}(a)\,l}\,+\,\mathcal{O}(k^{-1})\,.
\end{equation}
We can resubstitute this expression together with $\delta$ from \eqref{DeltaUV} and $C$ from  
\eqref{cosetOPEcoefficientC} into \eqref{lambdaIR}, which becomes
\begin{equation}
\lambda_{IR}=-\frac{\sqrt{{\rm dim}\,a}\,l}{\pi k}\left(1-\frac{\dcox}{k}+\mathcal{O}
(k^{-2})\right)\,.
\end{equation}
With this value of $\lambda$, and the expressions \eqref{cosetOPEcoefficientC} for $C$
and \eqref{cosetcoefficientD} for $D$, we obtain precisely the asymptotic
expressions of the $R_{ij}$ in \eqref{CosetR} from the general perturbative result 
\eqref{pertresult part1}, \eqref{pertresult part2}. 

For completeness let us also give the reflection coefficient:
\begin{equation}\label{CosetRT}
{\cal R}\,=\,\frac{l (l+\dcox) ((k+\dcox)^2-l(l+\dcox))}{(k+\dcox)^2 (k(k+2\dcox)-l(l+\dcox))}
\,=\,\frac{l(l+\dcox)}{k^2}\left(1\,-\,\frac{2\dcox}{k}\,+\,\mathcal{O}(k^{-2})\right)\,.
\end{equation}
Of course, ${\cal T}$ is given by $1-{\cal R}$.\\

Notice the limited information reflection and transmission provide for the task of actually fixing the
RG interface. As mentioned before, symmetry considerations lead us to search for the RG interface 
within the set given by a fusion product of the topological defect \eqref{GPtopologicaldefect} 
with a boundary state of the form \eqref{GPboundarystate}. However, all of these interfaces yield 
the same expressions \eqref{CosetR}, {\it i.e.} the same reflection and transmission, since these 
expressions only depend on the decomposition \eqref{Tdecomposition} of the energy-momentum tensors. In 
order to corroborate that \eqref{actualGPdefect} indeed is the RG interface, a calculation of actual
overlaps \eqref{GPoverlap} is needed. This has been done in perturbation theory in 
\cite{Poghosyan:2013qta,Poghosyan2} (see also \cite{Poghossian:2013fda}) in the case $a=su(2)$. Of course, for 
perturbative flows, {\it i.e.} large values of the level $k$, the claim that we have indeed correctly 
identified the RG interface follows from the relation \eqref{gandR}. The $g$ factor is the overlap of the two 
vacua, and with our formula \eqref{CosetRT} for ${\cal R}$ and \eqref{cosetOPEcoefficientC} for $c^{(1)}$, 
\eqref{gandR} leads to the condition
\begin{equation}\label{ourgevaluated}
g^2=\frac{(S_{R0}^{(k-l)}S_{R0}^{(k+l)}S_{R0}^{(l)})^2}{S_{00}^{(k-l)}S_{00}^{(k+l)}(S_{00}^{(l)})^2}
=1+\frac{{\rm dim}(a)\,l^2}{2k^2}-\frac{{\rm dim}(a)\,l^2\,\dcox}{k^3}+\mathcal{O}(k^{-4})\,.
\end{equation}
In appendix~\ref{appendix:gfunction} we use the 
general formula \eqref{Smatrixelement} for the relevant $S$ matrix elements together with the 
expansions \eqref{gexpansions} to check that order by order in $1/k$, this condition is met if 
and only if the RG defect is given by \eqref{actualGPdefect}, {\it i.e.} if all representation labels
$R^{(n)}=0$.

As pointed out in section~\ref{section:RT} we also observe that the relation \eqref{gandR} between the entropy 
and the reflection only holds perturbatively. Indeed, the exact result for the reflection coefficient
in \eqref{CosetRT} for ${\cal R}$ does not contain the same information about the 
algebra $a$ as the $S$ matrix elements do. As pointed out in appendix~\ref{appendix:gfunction}, 
the $\mathcal{O}(k^{-4})$ term in the expansion of the left-hand side of \eqref{ourgevaluated} depends 
nontrivially on the root system of the algebra, an information 
which does not appear in ${\cal R}$.

\section{Marginal perturbations}\label{section:limitcases}

We consider the results from section~\ref{section:RT} in the limit 
where the perturbation density is initially a marginal operator, $\delta=0$. In this case
the coefficients $C$ and $D$ in the beta function \eqref{betafunction} are universal, {\it i.e.}
independent of the choice of scheme.
The coefficient $C$ is therefore universally given by the OPE coefficient. If these coefficients 
vanish, the flow becomes exactly marginal (at least to the order in $\lambda$ considered here). 
If the solution of the beta function equation satisfies $C\lambda<0$, the perturbing field is 
marginally irrelevant. In this case, sufficiently small perturbations will drive the system 
back to the fixed point we started with. In the opposite case, where $C\lambda>0$, the 
perturbation becomes relevant, and generically the system will flow to a different fixed 
point in the IR.\\

Examples for the case where the perturbation is marginally irrelevant
are the coset model flows of section~\ref{section:GP defects} in the strict limit 
$k=\infty$. The limit system is a continous orbifold based on the algebra
$\hat{a}_l$ \cite{Gaberdiel:2011aa,Fredenhagen:2014kia,Fredenhagen:2012rb}. The 
limit of the perturbing field $\phi$ is a current-current deformation in the untwisted 
sector, and the Gaiotto-Poghosyan interfaces become the identity defect, consistent
with the fact that the reflection coefficient goes to zero. From the $k\rightarrow\infty$ 
limit of our formulae \eqref{cosetOPEcoefficientC}, \eqref{cosetcoefficientD} we see
that the field indeed represents a marginally irrelevant perturbation for any
initial (bare) value $\lambda<0$ \cite{Zamolodchikov:1991vg}. Notice that while in principle the
continuous orbifold limit can have deformations corresponding to exactly marginal operators in the 
untwisted sector (\cite{Forste:2003km}, see also \cite{Gaberdiel:2014cha}),
the limit of the coset flow perturbations will not be within this class.\\

A very simple instance for an exactly marginal perturbation is provided
by the free boson $X$ compactified on a circle of radius $R$. The deformation by the
operator $\phi=\partial X\bar{\partial}X$ corresponds to a change in the radius of the 
compactification. All conformal interfaces in the free boson theory were constructed explicitly
in \cite{Bachas:2001vj}. Here we are interested in the defect separating two compactification radii
$R_1$ and $R_2$, which becomes the identity interface in the limit $R_1=R_2$. In the folded 
picture the compactification is a rectangular torus, and this interface corresponds to the diagonal 
$U(1)$-preserving $D1$ brane that wraps once around both cycles. The boundary state has the 
form~\cite{Bachas:2001vj}
\begin{equation}\label{D1brane}
\braneket{D1,\vartheta}=g\prod_{n=1}^\infty e^{\frac{1}{n}S_{ij}^{(+)}a_{-n}^i\tilde{a}_{-n}^j}
\sum_{M,W\in\mathbb{Z}}\ket{M,W}\otimes\ket{-M,W}\,.
\end{equation}
Here $\tan\vartheta=R_2/R_1$ gives the angle of the brane, and $g=\sin^{-\frac{1}{2}}(2\vartheta)$
is the boundary entropy. The sum runs over a subset of $U(1)$ ground states of the torus compactification. 
In our notation each factor of a ground state corresponds to a torus cycle ({\it i.e.} we have
suppressed the right-moving labels), and the labels give momentum $M$ and winding $W$. The conformal 
dimension of such a ground state is \mbox{$\Delta_{M,W}=\tfrac{1}{2}(M R_1^{-1}+WR_1)^2+\tfrac{1}{2}(-M R_2^{-1}+WR_2)^2$}.
The $a_n^i$ and $\tilde{a}_n^j$ in \eqref{D1brane} denote $U(1)$ chiral and antichiral modes
of the cycle $i$, $j\in\{1,2\}$, respectively, with normalisation 
$[a^i_m,a^j_n]=m\delta^{ij}\delta_{m,-n}$. The coefficients $S^{(+)}_{ij}$ form the matrix
\begin{equation}
S^{(+)}=-\left(\begin{array}{cc}
\cos 2\vartheta & \sin 2\vartheta \\ \sin 2\vartheta & -\cos 2\vartheta
\end{array}\right)\,.
\end{equation}
If we map our cylindrical worldsheet to the plane, and fold the inside of the
unit disc to the outside, this boundary state is inserted along the unit circle.
The entries of the $R$ matrix \eqref{Rdefinition} in this setup are given by \cite{QRW}
\begin{equation}
R_{ij}=\frac{\bra{0}\tilde{L}_2^{(i)}L_2^{(j)}\braneket{D1}}{\langle 0\braneket{D1}}\,.
\end{equation}
Using standard commutation relations one quickly finds
\begin{equation}
R_{ij}=\frac{1}{2}\,S_{ij}^2\,,\quad {i.e.}\quad
\mathcal{R}=\cos^2(2\vartheta)\,,\;\mathcal{T}=\sin^2(2\vartheta)\,.
\end{equation}
These results agree with our perturbative formulae \eqref{pertresult part1}, 
\eqref{pertresult part2}, and \eqref{gandR}. Under the perturbation, the 
compactification radius $R_2$ changes from $R_2=R_1$ to
\begin{equation}
R_2=R_1e^{\pi\lambda}\,,
\end{equation}
such that 
\begin{equation}
\sin(2\vartheta)=\frac{2e^{\pi \lambda}}{1+e^{2\pi\lambda}}\,,\quad
\cos(2\vartheta)=\frac{1-e^{2\pi\lambda}}{1+e^{2\pi\lambda}}
\end{equation}
or
\begin{align}
R_{11}=R_{22}&=\frac{\pi}{2}\,\lambda^2\,+\,\mathcal{O}(\lambda^4)\,,\qquad
R_{12}=R_{21}=\frac{1}{2}\,-\,\frac{\pi}{2}\,\lambda^2\,+\,\mathcal{O}(\lambda^4)\,.
\end{align}
Notice in particular that in this case the $g$ factor and the reflection are indeed related by the simple 
formula
\begin{equation}\label{gandRfreeboson}
g^2\,=\,{\cal T}^{-\frac{1}{2}}=1+\tfrac{1}{2}{\cal R}+\tfrac{3}{8}{\cal R}^2 + \ldots\,.
\end{equation}
It would be interesting to understand if there is any simple relation between ${\cal R}$ and $g$ for exactly 
marginal RG defects also in the more general case.

Finally, we remark that for our radius-changing defects the transmission and reflection coefficients
have a rather explicit interpretation in terms of probabilities for transmission and reflection of oscillator 
modes \cite{QRW}. The transmission coefficient ${\cal T}$ is related to the determinant of the renormalised 
square of the Bogolyubov transformation connecting the modes \cite{Anatolyinprep}.

\section{Conclusion}\label{section:conclusion}
In this paper we derived a perturbative result for the reflection and transmission of energy and momentum
of conformal RG interfaces. The result for the entries of the matrix $R$, defined in \eqref{Rdefinition}, was 
given in \eqref{pertresult part1}, \eqref{pertresult part2}. The reflection coefficient \eqref{pertresultR} is 
the average of the reflections \eqref{pertresult part1} on the UV and the IR side of the interface. 
At least perturbatively, and in the examples we considered, the UV reflection is larger than the IR 
reflection in the case of relevant perturbations, and the two quantities are equal for marginal perturbations. 
From the results for the matrix $R$ we derived the perturbative relation \eqref{gandR} between the reflection 
coefficient and the entropy of the RG interface at the fixed point.

The perturbative result agrees with the one obtained from explicit RG interface solutions 
\cite{Gaiotto,Poghosyan2} of coset model flows \eqref{hoppingflow}. These interfaces, whose uniqueness we fixed 
by comparing with the perturbation theory result \eqref{gandR}, also show that the $g$ factor in general 
cannot depend on ${\cal R}$ (in combination with the central charge) alone. There are however cases where too 
few parameters are present in the theory, and $g$ can in fact be written as a function purely of ${\cal R}$ 
and $c$. A simple case where this occurs are exactly marginal deformations of the compactified free boson, 
leading to the relation \eqref{gandRfreeboson}. 

As a side remark we pointed out that the limit of the coset perturbations considered in 
section~\ref{section:GP defects} will not become exactly marginal deformations of the continuous orbifold in 
the limit of infinite level $k$. Rather, for $\lambda<0$ the perturbation is marginally irrelevant, while
for  $\lambda>0$ it is marginally relevant.\\

Obviously there remain many questions. 
For our coset models there also exist flows for $\lambda>0$, which lead to massive 
integrable models that were studied in \cite{Ahn:1990gn}. The RG flows of these 
perturbations are non-perturbative, and therefore we refrained from considering them 
here. Since the IR theories are trivial, we expect that the RG interfaces
are given by a particular boundary condition in the UV theory $M_{k,l}$,
{\it i.e.} a totally reflective RG interface. It would be interesting to understand if
this boundary condition has infinite entropy, or if one can associate a 
particular boundary condition at finite entropy to these massive flows.
It seems plausible that 
any RG flow with a non-trivial IR fixed point (to be more precise, any RG flow 
where the IR theory contains an energy-momentum tensor and is not a degeneration limit) corresponds 
to an interface with ${\cal T}\neq 0$.

For our coset model the case $a=su(N)$, $l=1$ leads to the $W_{k,N}$ theories, {\it i.e.} to
the bosonic version of the CFT duals to higher spin algebras on $AdS_3$ \cite{Gaberdiel:2014cha}. 
It would be interesting to have an interpretation of these RG interfaces in the $AdS_3$ bulk. 
In that context we would also like to understand the fusion of two RG interfaces \cite{Anatolyinprep}, 
and how reflection and transmission behave under it.

Another more general point that we have not touched
at all is whether the reflection of an RG interface is always minimal among the 
interfaces which preserve the same symmetry as the RG flow.  It seems plausible that this is 
related to the stability of the RG interface under boundary perturbations.
It would also be interesting to prove that the off-critical definition of ${\cal R}$ 
(or ${\cal T}$) indeed provides a quantity which behaves monotonically along (non-perturbative) relevant RG 
flows.

Finally, while we have focused solely on the reflection and transmission of Virasoro
modes, studying the reflection and transmission of conserved currents 
might lead to a more refined picture of the set of RG interfaces. 

\section*{Acknowledgements}
We thank Anatoly Konechny, Charles Melby-Thompson, and Enrico Brehm for useful discussions.
This work was supported in part by the DFG Transregio grant TRR33.


\appendix
\renewcommand{\theequation}{\Alph{section}.\arabic{equation}}

\setcounter{equation}{0}
\section{Details of perturbative calculations}\label{appendix:perturbativecalculations}
In this appendix we collect some details of the computations in section~\ref{section:RT}.
In general, there are several ways to do the integrals; we present one which seems 
to be convenient to us.

\subsection{Computation of $R_{11}^{(3)}$}
For $R_{11}^{(3)}$ we have to do the integral in the last line of~\eqref{R11order3}.
The first part involves the integral~\eqref{R11order3firstpart}. We have
\begin{equation}\label{R11firstpart}
\int d^2z_1d^2z_2\left|\frac{1}{z_{21}}\right|^2=2\int d^2z_1 \int d^2\xi\,|1-\xi|^{-2}\,.
\end{equation}
On the right-hand side we can map the $\xi$ integral to the upper half plane
with \mbox{$\eta=i(1-\xi)/(1+\xi)$}, and reflect the part outside the unit disc
by $\eta\rightarrow 1/\bar{\eta}$:
\begin{equation}\label{basic1}
\int d^2\xi\,|1-\xi|^{-2}=\int_{\mathbb{D}^+}d^2\eta|1-i\eta|^{-2}(1+|\eta|^{-2})\,.
\end{equation}
The integration region $\mathbb{D}^+$ denotes the upper half-disc. Consider
the part on the right-hand side of \eqref{basic1} which has no divergence. By 
Stokes' theorem we have
\begin{equation}\label{R11firstpartplane}
\int_{\mathbb{D}^+}d^2\eta|1-i\eta|^{-2}=-\frac{1}{2}\oint_{\partial\mathbb{D}^+}d\eta\,
\frac{\log(1+i\bar{\eta})}{1-i\eta}\,.
\end{equation}
The boundary consists of two pieces such that
\begin{align}
-\frac{1}{2}\oint_{\partial\mathbb{D}^+}d\eta\,\frac{\log(1+i\bar{\eta})}{1-i\eta}=
-\frac{1}{2}\int_{-1}^1d\eta\,\frac{\log(1+i\eta)}{1-i\eta}
-\frac{1}{2}\int_{\gamma}d\eta\,\frac{\log(1+i/\eta)}{1-i\eta}\,,
\end{align}
where $\gamma$ is the counterclockwisely oriented upper half-circle.
One can now expand the integrands on the right-hand side in small 
$\eta$ and resum, or use the elementary integals
\begin{align}
\frac{1}{i}\int d\eta\,\frac{\log(1+i\eta)}{1-i\eta}&=\log(1+i\eta)\log(\tfrac{1-i\eta}{2})
+{\rm Li}_2(\tfrac{1+i\eta}{2})\,,\nonumber\\
\frac{1}{i}\int d\eta\,\frac{\log(\eta+i)}{1-i\eta}&=\frac{1}{2}\log(i+\eta)^2\,,\\
\frac{1}{i}\int d\eta\,\frac{\log(\eta)}{1-i\eta}&=\log(\eta)\log(1-i\eta)+
{\rm Li}_2(i\eta)\,.\nonumber
\end{align}
In the end one finds the results
\begin{align}
-\frac{1}{2}\int_{-1}^1d\eta\,\frac{\log(1+i\eta)}{1-i\eta}&={\rm Cat}-
\frac{3\pi}{8}\log 2\,,\nonumber\\
-\frac{1}{2}\int_{\gamma}d\eta\,\frac{\log(1+i/\eta)}{1-i\eta}&={\rm Cat}-
\frac{\pi}{8}\log 2\,,
\end{align}
such that
\begin{equation}\label{basic2}
\int_{\mathbb{D}^+}d^2\eta|1-i\eta|^{-2}=2{\rm Cat}-\frac{\pi}{2}\log 2\;
\approx\; 0.743138\,.
\end{equation}
In the last equations, Cat is Catalan's constant
\begin{equation}\label{Catalan}
{\rm Cat} = \sum_{k=0}^\infty \frac{(-1)^k}{(2 k+1)^2}=0.915966\,.
\end{equation}
In order to calculate the divergent part in \eqref{basic1} we expand
\begin{align}\label{expansion}
\int d^2\eta|1-i\eta|^{-2}|\eta|^{-2}&=\sum_{m,n=0}^\infty\int dr %
r^{m+n-1}i^{m-n}\int_0^\pi d\phi\, e^{i\phi(m-n)}\nonumber\\
&= \sum_{m,n=0}^\infty\int dr r^{(m-1)+(n-1)+1}i^{(m-1)-(n-1)}\int_0^\pi d\phi\, %
e^{i\phi((m-1)-(n-1))}\\
&=\int dr r^{-1}\pi+\sum_{m,n=0}^\infty\int dr r^{m+n+1}i^{m-n}\int_0^\pi d\phi\, %
e^{i\phi(m-n)}\nonumber\\
&\quad+\sum_{m=1}^\infty \int dr r^{m-1}\int_0^\pi d\phi %
\left(i^me^{im\phi}+i^{-m}e^{-im\phi}\right)\,.\nonumber
\end{align} 
In the penultimate line, the first integral contains the logarithmically divergent 
term. The double sum is obtained from shifting the labels $m$ or $n$, and it is in 
fact nothing but the expansion of \eqref{R11firstpartplane},
\begin{equation}
\int_{\mathbb{D}^+}d^2\eta|1-i\eta|^{-2}=\sum_{m,n=0}^\infty\int dr r^{m+n+1}i^{m-n}%
\int_0^\pi d\phi\, e^{i\phi(m-n)}\,.
\end{equation}
In the last line in \eqref{expansion} there are the contributions from $n=0,\,m>0$ 
and $m=0,\,n>0$. All even terms vanish. Writing $m=2k+1$, the integral is easily done:
\begin{align}
&\sum_{m=1}^\infty \int dr r^{m-1}\int_0^\pi d\phi \left(i^me^{im\phi}+%
i^{-m}e^{-im\phi}\right)=\nonumber\\
&\quad \sum_{k=0}^\infty2\re\left[\frac{2i^{2k+2}}{(2k+1)^2}\right] \;=\;%
-4\sum_{k=0}^\infty \frac{(-1)^k}{(2k+1)^2}\;=-4{\rm Cat}\,.
\end{align}
Combining the results for the divergent and the non-divergent part of \eqref{basic1}
we therefore have
\begin{equation}\label{basic3}
\int d^2\xi|1-\xi|^{-2}=\int dr\frac{\pi}{r}+2\int_{\mathbb{D}^+}d^2%
\eta|1-i\eta|^{-2}-4{\rm Cat}=\int dr\frac{\pi}{r}-\pi\log 2\,.
\end{equation}
In the $\eta$ coordinates, the cut-off is $\epsilon/2+\mathcal{O}(\epsilon^2)$, from which we
conclude that
\begin{equation}\label{divpart1}
\int \frac{\pi\,dr}{r}=-\pi\log(\frac{\epsilon}{2})\, +\, \mathcal{O}\left(\epsilon\right)\,.
\end{equation}
There is therefore no finite contribution (of $\mathcal{O}(\epsilon^0)$) in \eqref{basic3}.
Moreover, the subsequent integral over $z$ in \eqref{R11firstpart} only gives a factor,
and thus \eqref{R11firstpart} is entirely cancelled by the counterterm in our scheme.\\

The second part of \eqref{R11order3} is given by the integral on the left-hand side
of \eqref{R11order3secondpart},
\begin{equation}
\int d^2z_1d^2z_2d^2z_3\frac{1}{z_{12}\bar{z}_{23}}\,.
\end{equation}
There is no cut-off necessary to compute this integral. By symmetry we can take 
$|z_1|\leq |z_3|$ at the cost of an overall factor of 2. Then there remain three regions:
\begin{itemize}
\item[a)] $|z_2|\leq |z_1|\leq |z_3|$\\
Performing the change of coordinates $z_1=\xi z_3$, $z_2=\eta z_1=\eta\xi z_3$ we obtain
\begin{align}\label{caseA}
2\int^* d^2z_1d^2z_2d^2z_3\frac{1}{z_{12}\bar{z}_{23}}&=2\int d^2z_3d^2\xi d^2\eta\,
\frac{|z_3|^2\bar{\xi}}{(1-\eta)(\bar{\eta}\bar{\xi}-1)}\nonumber\\
&=-2\int d^2z_3d^2\xi d^2\eta\,|z_3|^2\sum_{m,n=0}^\infty\eta^m\bar{\eta}^n\bar{\xi}^{n+1} =0\,.
\end{align}
In the integral on the left, the star is there to remind us of the special
condition on $z_1,\,z_2$ and $z_3$. In the last line, the integral over $\xi$ 
vanishes by the angular integration. 
\item[b)] $|z_1|\leq |z_2|\leq |z_3|$\\
Analogously to case a) we change coordinates $z_2=\xi z_3$, $z_1=\eta z_2=\eta\xi z_3$ and find
\begin{align}\label{caseB}
2\int^* d^2z_1d^2z_2d^2z_3\frac{1}{z_{12}\bar{z}_{23}}&=2\int d^2z_3d^2\xi d^2\eta\,
\frac{|z_3|^2\bar{\xi}}{(1-\eta)(1-\bar{\xi})}=0\,,
\end{align}
again due to the angular integration in the $\xi$ integral.
\item[c)] $|z_1|\leq |z_3|\leq |z_2|$\\
Proceeding as before we change $z_3=\xi z_2$, $z_1=\eta z_3=\eta\xi z_2$. Then we have
\begin{align}\label{caseC}
2\int^* d^2z_1d^2z_2d^2z_3\frac{1}{z_{12}\bar{z}_{23}}&=2\int d^2z_3d^2\xi d^2\eta\,
\frac{|z_2|^2|\xi|^2}{(\eta\xi-1)(1-\bar{\xi})}\nonumber\\
&=-2\int d^2z_3d^2\xi d^2\eta\,\sum_{m,n=0}^\infty|z_2|^2|\xi|^2\eta^m\xi^m\bar{\xi}^n\nonumber\\
&=-\frac{\pi^3}{2}\,.
\end{align}
\end{itemize}
Combining \eqref{caseA}, \eqref{caseB}, and \eqref{caseC}, the result \eqref{R11order3secondpart}
follows.

\subsection{Computation of $R_{22}^{(3)}$}
For $R_{22}^{(3)}$ we need to compute the right-hand side of \eqref{longint}.
In this calculation we have to keep track of the transforming cut-offs. We must 
perform the integration over the coordinates $\eta,\,\xi,$ and $z_3$ in this order. 
The cut-offs in $\eta$ and $\xi$ are $\epsilon_\eta=\tfrac{\epsilon_\xi}{|\xi|}$, 
and $\epsilon_\xi=\tfrac{\epsilon_{z_3}}{|z_3|}$, respectively. Notice that the 
integrals over $z_3$ need to be cut off for $z_3\rightarrow0$. There the cut-off is given 
by the IR cut-off on the cylinder. Eventually we can simply work with an otherwise 
unspecified cut-off $\epsilon_{z_3}$, and collect the finite contributions 
${\cal O}(\epsilon_{z_3}^0)$. In the following, we use the symbol $\sim$ to refer to 
this part of the integral. We will repeatedly use the following identities:
\begin{align}\label{listofintegrals1}
\int d^2 x|x|^{-2k-2}&=\frac{\pi}{k\epsilon_x^{2k}}\,-\,\frac{\pi}{k}\,,\nonumber\\
\int d^2x|x|^{-2k}|1-x|^{-2}&=\sum_{m=0}^{k-2}\frac{\pi\epsilon_x^{-(2k-2m-2)}}{k-m-1}\,-\,3\pi\log\epsilon_x
\,-\,\pi H_{k-1}\,,\nonumber\\
\int d^2\eta|\eta(1-\xi\eta)|^{-2}&=-2\pi\log\epsilon_\eta\,-\,\pi\log(1-|\xi|^2)\,,\nonumber\\
\int d^2\eta|\eta|^{-2}(1-\eta)^{-1}(1-\bar{\xi}\bar{\eta})^{-1}&=
-2\pi\log\epsilon_\eta\,-\,\pi\log(1-\bar{\xi})\,,\\
-\int d^2\xi |\xi|^{-4}\log(1-|\xi|^2)&=-2\pi\log\epsilon_\xi\,+\,\pi\,,\qquad
\int d^2\xi|\xi|^{-4}\log\bar{\xi}=0\,,\nonumber\\
\int d^2x|x|^{-6}\log|x|&=\frac{\pi}{8\epsilon_x^4}+\frac{\pi\log\epsilon_x}{2\epsilon_x^4}-\frac{\pi}{8}\,,
\nonumber\\
\int d^2x|x|^{-4}\log|x|&=\frac{\pi}{2\epsilon_x^2}+\frac{\pi\log\epsilon_x}{\epsilon_x^2}-\frac{\pi}{2}\,.
\nonumber
\end{align}
Here $k\in\mathbb{N}$, $H_k=\sum_{n=1}^kn^{-1}$ is the $n$th harmonic number,
end empty sums are zero.\\
Consider now the right-hand side of \eqref{longint} summand by summand:
\begin{enumerate}
\item In the first summand the integrand is $1/|z_3^{6}\,\xi^{4}\,\eta^{4}\,(1-\xi)^2|$. Use
\begin{equation}
\int d^2\eta|\eta|^{-4}=\frac{\pi}{\epsilon_\eta^2}-\pi=\frac{\pi}{\epsilon_{\xi}^2}|\xi|^2-\pi\,,
\end{equation}
such that the subsequent $\xi$ integral becomes
\begin{equation}
\pi\int d^2\xi\frac{\frac{\pi}{\epsilon_{\xi}^2}|\xi|^2-\pi}{|\xi|^4|1-\xi|^2}=-\frac{\pi^2}{\epsilon_\xi^2}
(1+3\log\epsilon_\xi)+3\pi^2\log\epsilon_\xi+\pi^2\,.
\end{equation}
Using \eqref{listofintegrals1} we get a finite contribution
\begin{equation}\label{finitecontribution1}
\int d^2z_3d^2\xi d^2\eta\frac{|z_3|^{-6}|\xi\eta|^{-4}}{|1-\xi|^2}\sim -\frac{\pi^3}{8}\,.
\end{equation}

\item In the second summand, the integrand is $1/|z_3^{6}\,\xi^{4}\,\eta^{2}\,(1-\eta)^2|$. Use
\begin{equation}
\int d^2\eta|\eta(1-\eta)|^{-2}=-3\pi\log\epsilon_\eta=-3\pi\log\epsilon_\xi+3\pi\log|\xi|\,,
\end{equation}
such that
\begin{equation}
-3\pi\int d^2\xi|\xi|^{-4}\left(\log\epsilon_\xi-\log|\xi|\right)=
\frac{3\pi^2}{2\epsilon_\xi^2}+3\pi^2\log\epsilon_\xi-\frac{3\pi^2}{2}\,.
\end{equation}
Using \eqref{listofintegrals1} again one finds
\begin{equation}\label{finitecontribution2}
\int d^2z_3d^2\xi d^2\eta\frac{|z_3|^{-6}|\xi|^{-4}|\eta|^{-2}}{|1-\eta|^2}\sim\frac{9\pi^3}{8}\,.
\end{equation}

\item For the third summand, the integrand is $1/|z_3^{6}\,\xi^{4}\,\eta^{2}\,(1-\xi\eta)^2|$. 
The $\eta$ integral is
\begin{equation}
\int d^2\eta\frac{|\eta|^{-2}}{|1-\xi\eta|}=\int d^2\eta\sum_{m=0}^\infty|\eta|^{2m-2}|\xi|^{2m}
=-2\pi\log\epsilon_\eta-\pi\log(1-|\xi|^2)\,.
\end{equation}
Integrating the result with $\epsilon_\eta=\epsilon_\xi|\xi|$ against $\int d^2\xi|\xi|^{-4}$,
we obtain by means of the integrals \eqref{listofintegrals1}
\begin{equation}
\int d^2\xi d^2\eta\frac{|\xi|^{-4}|\eta|^{-2}}{|1-\xi\eta|^2}=\frac{\pi^2}{\epsilon_{\xi}^2}=
\frac{\pi^2}{\epsilon^2}|z_3|^2\,.
\end{equation}
The following integration over $z_3$ does not change the finite part of this result, 
such that 
\begin{equation}
\int d^2z_3d^2\xi　d^2\eta\frac{|z_3|^{-6}|\xi|^{-4}|\eta|^{-2}}{|1-\xi\eta|^2}\sim 0\,.
\end{equation}

\item The fourth and fifth summands in \eqref{longint} do not contribute any finite
quantities because of the angular integration in $\eta$. 

\item Expanding the integrand $2|z_3^{-6}\,\xi^{-4}\,\eta^{-2}|(1-\eta)^{-1}(1-\bar{\xi}\bar{\eta})^{-1}$ 
of the last summand in $\eta$ we obtain for the $\eta$ integral
\begin{equation}
-2\int d^2\eta\frac{|\eta|^{-2}}{(1-\eta)(1-\bar{\xi}\bar{\eta})}=4\pi \log\epsilon_\eta 
+2\pi\log(1-\bar{\xi})\,.
\end{equation} 
The logarithmic term drops out in the angular integration over the $\xi$ coordinate,
but the first term leaves us with the contribution
\begin{equation}
-2\int d^2\xi d^2\eta\frac{|\xi|^{-4}|\eta|^{-2}}{(1-\eta)(1-\bar{\xi}\bar{\eta})}=
-\frac{2\pi^2}{\epsilon_\xi^2}-4\pi\log\epsilon_\xi+2\pi^2\,.
\end{equation}
The $z_3$ integration then shows that the contribution from this summand is
\begin{equation}\label{finitecontribution3}
-2\int d^2z_3d^2\xi d^2\eta\frac{|z_3|^{-6}|\xi|^{-4}|\eta|^{-2}}{(1-\eta)(1-\bar{\xi}\bar{\eta})}\sim
-\frac{3\pi^2}{2}\,.
\end{equation}
\end{enumerate}
Combining the non-vanishing contributions \eqref{finitecontribution1}, 
\eqref{finitecontribution2}, and  \eqref{finitecontribution3} gives the result 
\eqref{R22order3result} in our renormalisation scheme.

\setcounter{equation}{0}
\section{Expansion of $g$ factors}\label{appendix:gfunction}
In this appendix we collect some formulae needed in the perturbative argument that \eqref{actualGPdefect}
is the actual RG interface among the class $D\braneket{B}$ given by \eqref{GPboundarystate},
\eqref{GPtopologicaldefect}. The formula for the modular $S$ matrix elements can be found in the standard
literature (see {\it e.g.} \cite{DiFrancesco}). We will only need an expression for $S$ matrix
elements of the form $S_{r0}$, for which the general expressions simplify to
\begin{equation}\label{Smatrixelement}
S_{R0}^{(k)}=|\det((\alpha^{\!\vee}_i)_j)|^{-\frac{1}{2}}(k+\dcox)^{-\frac{r}{2}}
\prod_{\alpha\in\Delta_+}2\sin\left(\frac{\pi(\alpha,R+\rho)}{k+\dcox}\right)\,.
\end{equation}
In this expression, $R$ is an (affine) representation of $\hat{a}_{k}$, $\alpha^{\!\vee}_i$
denote the coroots of the horizontal algebra $a$ for $i=1,\ldots,r$ with $r$ the rank
of $a$, $\dcox$ is the dual Coxeter number of $a$, $\Delta_+$ denotes the set of 
positive roots of $a$, and $\rho$ is the Weyl vector.\\
The $g$ factor of all branes $D\braneket{B}$ in section~\ref{section:RT} is given by
\begin{equation}\label{generalg}
g=\frac{S^{(k-l)}_{R0}S^{(k+l)}_{R0}}{\sqrt{S^{(k-l)}_{00}S^{(k+l)}_{00}}}
\frac{S^{(l)}_{R0}}{S^{(k)}_{R0}S^{(l)}_{00}}\,,
\end{equation}
where we dropped the level labels on the representations. In this expression,
overal factors in the $S$ matrices which do not depend on the level drop out, and we are
left with factors of the following type:

\begin{align}\label{gexpansions}
F_1&=\frac{(k+\dcox)^\frac{r}{2}}{(k-l+\dcox)^{\frac{r}{4}}(k+l+\dcox)^{\frac{r}{2}}}
=1+\frac{l^2r}{4k^2}-\frac{\dcox l^2 r}{2k^3}+\mathcal{O}(k^{-4})\,,\nonumber\\[5pt]
F_2(\alpha)&=\frac{s^{(k-l)}(R)s^{(k+l)}(R)}{\sqrt{s^{(k-l)}(0)s^{(k+l)}(0)}}
\frac{s^{(l)}(R)}{s^{(k)}(R)s^{(l)}(0)}\\
&\quad=\frac{s^{(l)}(R)}{s^{(l)}(0)}\sqrt{\frac{\iota(R^{(k-l)})\iota(R^{(k+l)})}{\iota(R^{(k)})\iota(0)}}
\bigg(1+\frac{1}{6k^2}F_{22} - \frac{1}{3k^3}F_{23} + \mathcal{O}(k^{-4})\bigg)\,.\nonumber
\end{align}
In \eqref{gexpansions} we have defined the notation
\begin{align}
s^{(n)}(R)&=\sin\left(\frac{\pi (\alpha,R^{(n)}+\rho)}{n+\dcox}\right)\,,\qquad\qquad
\iota(R)=(\alpha,R+\rho)^2\,,\nonumber\\
F_{22}&=3l^2+\pi^2\big(\iota(0)+\iota(R^{(k)})-\iota(R^{(k-l)})-\iota(R^{(k+l)})\big)\,,\\[5pt]
F_{23}&=\dcox F_{22}+l\pi^2\big(\iota(R^{(k-l)})-\iota(R^{(k+l)})\big)\,.\nonumber
\end{align}
With these expressions we obtain an expansion
\begin{equation}
g^2 = F_1^2\left(\prod_{\alpha\in\Delta_+}F_2(\alpha)\right)^2
\end{equation}
from \eqref{generalg}, which must agree order by order in $1/k$ with the expansion \eqref{ourgevaluated}.
This in turn fixes all representations $R$ of the boundary state corresponding to the RG interface
to be trivial. Indeed, already in the $k\rightarrow \infty$ limit, the requirement that the RG interface
becomes the identity defect forces $R^{(l)}=0$. In the orders $1/k^2$ and $1/k^3$, the fact that there
is no factor of $\pi^2$ appearing in \eqref{ourgevaluated} imposes the necessary further
restrictions on the other representations $R^{(k\pm l)}$ and $R^{(k)}$. Although the factors
$F_2(\alpha)$ in the end still contain contributions $\iota(0)=(\alpha,\rho)^2$, these drop out in the 
expansion to the order we are considering, leaving only the information on the number $|\Delta_+|$ of positive 
roots. In the end one obtains
\begin{equation}
g^2=1+\frac{l^2(r+2|\Delta_+|)}{2k^2}-\frac{\dcox l^2(r+2|\Delta_+|)}{k^3}+\mathcal{O}(k^{-4})\,.
\end{equation}
By the standard Chevalley decomposition $r+2|\Delta_+|={\rm dim}(a)$, such that the expression indeed 
reproduces \eqref{ourgevaluated}. We remark that in the next order $1/k^4$,
the expansion starts to depend on the products $\pi(\alpha,\rho)$. 
From the perturbation theory point of view, this reflects the fact that intermediate channels in 
the four-point function begin to play an important role in the computation of the coefficient of $1/k^4$ 
\cite{Konechny:2014opa}. To the next higher order in ${\cal R}$, the relation \eqref{gandR} will therefore 
depend on more details of the CFT than just the reflection and the central charges.

\end{document}